\documentclass[journal]{IEEEtran}
\usepackage[latin1]{inputenc}
\usepackage{amsmath}
\usepackage{amssymb}
\usepackage{graphicx}
\usepackage{wasysym}
\usepackage[unicode=true,
 bookmarks=false,
 breaklinks=false,pdfborder={0 0 1},backref=section,colorlinks=false]
 {hyperref}
\hypersetup{
 hidelinks}

\makeatletter
\usepackage{amsfonts}
\usepackage{multirow}
\providecommand{\U}[1]{\protect \rule{.1in}{.1in}}
\makeatother

\begin{document}
\title{State-Compensation-Linearization-Based Stability Margin Analysis for
a Class of Nonlinear Systems: A Data-Driven Method}
\author{Jinrui Ren, Quan Quan \thanks{This work was supported by {*}{*}{*}. \textsl{(Corresponding authors:
Quan Quan).}\protect \\
Jinrui Ren is with the School of Automation, Xi\textquoteright an
University of Posts \& Telecommunications, Shaanxi, Xi\textquoteright an,
China; Quan Quan is with the School of Automation Science and Electrical
Engineering, Beihang University, Beijing 100191, China. (e-mail: renjinrui@xupt.edu.cn;
qq\_buaa@buaa.edu.cn).}}
\maketitle
\begin{abstract}
The classical stability margin analysis based on the linearized model
is widely used in practice even in nonlinear systems. Although linear
analysis techniques are relatively standard and have simple implementation
structures, they are prone to misbehavior and failure when the system
is performing an off-nominal operation. To avoid the drawbacks and
exploit the advantages of linear analysis methods and frequency-domain
stability margin analysis while tackling system nonlinearity, a state-compensation-linearization-based
stability margin analysis method is studied in the paper. Based on
the state-compensation-linearization-based stabilizing control, the
definition and measurement of the stability margin are given. The
$\mathcal{L_{\text{\ensuremath{2}}}}$ gain margin and $\mathcal{L_{\text{\ensuremath{2}}}}$
time-delay margin for the closed-loop nonlinear system with state-compensation-linearization-based
stabilizing control are defined and derived approximatively by the
small-gain theorem in theory. The stability margin measurement can
be carried out by the frequency-sweep method in practice. The proposed
method is a data-driven method for obtaining the stability margin
of nonlinear systems, which is practical and can be applied to practical
systems directly. Finally, three numerical examples are given to illustrate
the effectiveness of the proposed method.
\end{abstract}


\section{Introduction}

With the development of society, more and more fields demand high
safety and reliability. However, uncertainties are ubiquitous and
inevitable for almost all practical systems \cite{zhou1996robust}.
In practice, it is not enough to just know whether a system is stable
or not. It needs to be further analyzed how stable and robust a system
is. As a quantitative indicator of the systems' relative stability
and robustness, stability margin (SM) \cite{dorf2011modern,lichter2009flight}
is widely used in control engineering. Uncertainties, such as delay
or unmodeled higher-order dynamics, can be intuitively captured by
SM, which can be further adopted as a safety metric for danger monitoring,
controller tuning \cite{ho1996performance,tan2012advances}, filter
and compensator design \cite{ren2016initial}, etc. Thus, SM is important
in the analysis and design of control systems.

There has been much research on SM. For single-input single-output
(SISO) linear time-invariant (LTI) systems, the classic SM, namely
the gain margin (GM) and phase margin (PM), is well defined and understood
\cite{horowitz2013synthesis}. It is a class of classic frequency-domain
SM, which is straightforward, intuitive, and has clear physical meaning.
Thus, the classic SM is popular and widely adopted by control engineers.
Reference \cite{zhou1996robust} discussed robust SM in the field
of robust control. $\mu$-analysis is popular among multivariable
systems, in which the singular-value-based stability margin is defined
as a robust SM \cite{kim2017stability}. In order to deal with the
modeling uncertainties described by coprime factor perturbations,
the gap metric \cite{french2008adaptive} and its variant, the $\nu$-gap
\cite{konghuayrob2017low}, provide quantitative measures of robustness
stability. Time-delay SMs for the control of processes with uncertain
delays were investigated in \cite{bozorg2006control}. However, these
robust SMs are mostly applicable to linear systems. 

For nonlinear systems, SM analysis becomes more complex and difficult.
The classic frequency-domain SM is inapplicable because frequency-domain
analysis and synthesis cannot be directly carried out in the presence
of nonlinearity. The circle criterion, Popov's criterion, and Kalman-Yakubovich-Popov
(KYP) lemma are used for the absolute stability analysis of a nonlinear
system in the frequency domain when it is a feedback connection of
a linear dynamical system and a nonlinear element satisfying the sector
condition \cite{todeschini2015adaptive,li2015frequency,khalil2002nonlinear},
in which Lyapunov functional approach is often used and further formulated
in the form of linear matrix inequalities (LMIs). However, few SMs
have been defined therein. Based on singular perturbation analysis,
the singular perturbation margin and generalized gain margin were
proposed in \cite{yang2015singular}. These metrics are bijective
counterparts of GM and PM. Besides, gain, sector, and disk margins
\cite{chellaboina2000stability,ahn2013new} are studied in terms of
different stability specifications, such as Lyapunov stability, $\mathcal{H}_{\infty}$
stability, or input-to-state stability. Although many SM definitions
have been proposed for nonlinear systems, most SMs for nonlinear systems
require time-domain definitions or calculations, which are complex
and not easy to apply in practice.

In practice, it is a common and mature practice to utilize linear
control techniques at discrete operating points within the operating
envelope or along a nominal trajectory, and then classic frequency-domain
SM is used to certify the designed controllers at the discrete points\cite{lichter2009flight,regan2008flight}.
The drawback of this approach is that the obtained SM may be unreliable
when the assumption of benign nonlinearity fails in operation. Another
common choice is experimental methods, for example, Monte Carlo simulations.
However, most experimental methods are time-consuming and costly.
What is worse, the worst situation may not occur, which leads to an
inaccurate estimate of SM. 

On the whole, while SM is of importance in control design practice,
it is hard to find the SM of a large number of nonlinear systems,
which motivates us to carry out this research. In this paper, a state-compensation-linearization-based
stabilizing control from \cite{quan2024state} is utilized for a class
of nonlinear systems. This stabilizing control can put all the uncertainties
in the primary linear system and compensate for all the nonlinearity
in the secondary nonlinear system, which can make the following SM
analysis for nonlinear systems easier. After state-compensation-linearization-based
stabilizing control, SM analysis and measurement are next focused.
Thanks to the developed control framework, frequency-domain SM analysis
methods are applicable to the resulting nonlinear systems. The SM
analysis can be divided into two steps. First, the SM for the primary
system (an MIMO linear system) is derived by using the small-gain
theorem. We have proposed a simple and practical data-driven stability
margin analysis method for linear multivariable systems in \cite{https://doi.org/10.1002/rnc.7413}.
Second, the SM for the whole system (an MIMO nonlinear system) is
derived based on the small-gain theorem. Both SMs can be obtained
by frequency-sweep experimental methods, which are preferable to calculation
methods in practice. Based on these, the final SM can be determined.
In the end, the whole procedure of the SCLC and SM analysis is summarized. 

The main contributions of this paper lie in: New SMs are defined based
on SCLC, and the corresponding measurement can be implemented by real
experiments (e.g., frequency-sweep experiments), not only numerical
calculation. The SM measurement by frequency-sweep experiments is
data-driven, which approaches reality more. Owing to SCLC, frequency-domain
SM analysis is extended smoothly to nonlinear systems. Furthermore,
a complete design and analysis procedure in terms of robustness and
stability for a class of nonlinear systems is presented, which is
practical and can be applied to practical systems directly. 

The paper is organized as follows. \textit{Section \ref{sec:Problem}}
gives the problem formulation including the considered system model
and the objectives. Then, how to design a state-compensation-linearization-based
stabilizing controller is recalled briefly in \textit{Section \ref{Sec:Control}}.
Next, the state-compensation-linearization-based SM analysis and measurement
are given in \textit{\ref{sec:SM}}. In \textit{Section \ref{sec:Simulations}},
simulations are performed to demonstrate the effectiveness and practicability
of the proposed method. In the end, \textit{Section \ref{Sec:Conclusions}}
presents the conclusions.

\section{Problem Formulation}

\label{sec:Problem}

Consider a class of MIMO nonlinear systems as

\begin{equation}
\dot{\mathbf{x}}(t)=\mathbf{A\mathbf{x}}(t)+\mathbf{f}(\mathbf{x}(t))+\mathbf{B}\mathbf{\boldsymbol{\mu}}(t),\mathbf{x}(0)=\mathbf{x}_{0}.\label{eq:Sys0}
\end{equation}
Here $\mathbf{A}\in\mathbb{R}^{n\times n}$ is a stable constant matrix,
$\mathbf{B}\in\mathbb{R}^{n\times m}$ is a constant matrix, $\mathbf{f}(\cdot):\mathbb{R}^{n}\rightarrow\mathbb{R}^{n}$
is a nonlinear function vector with $\mathbf{f}(\mathbf{0})=\mathbf{0}$,
$\mathbf{x}(t)\in\mathbb{R}^{n}$ is the state vector, and $\mathbf{\boldsymbol{\mu}}(t)\in\mathbb{R}^{m}$
is the actual control input.

We define the gain and time-delay margins by putting a complex diagonal
multiplicative perturbation $\mathbf{I}_{{m}}+\Delta(s)$ at the plant
input, namely

\begin{equation}
\mathbf{\boldsymbol{\mu}}(s)=(\mathbf{I}_{{m}}+\Delta(s))\mathbf{u}(s)\label{eq:UncertainyIn}
\end{equation}
where $\mathbf{u}(t)\in\mathbb{R}^{m}$ is the designed control input,
$\mathbf{I}_{{m}}\in\mathbb{R}^{m\times m}$ is an identity matrix,
and $\Delta(s)\in\mathbb{R}^{m\times m}$ is the artificial unmodeled
dynamics (it does not exist but is introduced only for the stability
margin analysis, and can be regarded as a ``stability margin gauge''
\cite{yang2015singular}). For system (\ref{eq:Sys0}), the following
assumptions are made. 

\textbf{Assumption 1.} The pair $(\mathbf{A},\mathbf{B})$ is controllable
\footnote{With such an assumption, $\mathbf{A}\in\mathbb{R}^{n\times n}$ is
stable without loss of generality.}. 

\textbf{Assumption 2.} The system state is measurable.

\textbf{Remark 1.} Noteworthy, there are no restrictions on the function
$\mathbf{f}(\mathbf{x}(t))$. For general systems $\mathbf{x}(t)=\mathbf{F}(x)+\mathbf{B}\mathbf{\mu}(t)$
which do not satisfy the form (\ref{eq:Sys0}), by letting $\mathbf{A}=\left.(\partial(\mathbf{F}(\mathbf{x}))/(\partial\mathbf{x}))\right|_{\mathbf{x}=\mathbf{0}}$,
$\mathbf{f}(\mathbf{x})=\mathbf{F}(\mathbf{x})-\mathbf{A}\mathbf{x}$,
they can be transformed into the considered systems in the form of
(\ref{eq:Sys0}). Therefore, the form (\ref{eq:Sys0}) is more general.

\textbf{Objective 1.} Under \emph{Assumptions 1-2}, the first objective
is to propose a controller $\mathbf{u}(t)=\mathbf{C}(\mathbf{x}(t))$
to make $\mathbf{x}\in\mathcal{L_{\text{\ensuremath{2}}}^{\text{\ensuremath{n}}}}$
under any initial condition $\mathbf{x}_{0}$ when $\Delta(s)=\mathbf{0}$.

Before introducing the next objective, the following stability margin
definition is introduced first.

\textbf{Definition 1.} Given any $\Delta(s)\equiv\text{diag}\left(\gamma_{1},\gamma_{2},\ldots,\gamma_{m}\right)$,
$\left|\gamma_{i}\right|\leq\gamma_{\text{max}}$, $i=1,2,\ldots m$,
if the state of the closed-loop system composed by (\ref{eq:Sys0}),
(\ref{eq:UncertainyIn}) and $\mathbf{u}(t)=\mathbf{C}(\mathbf{x}(t))$
satisfies $\mathbf{x}\in\mathcal{L_{\text{\ensuremath{2}}}^{\text{\ensuremath{n}}}}$
for any initial condition $\mathbf{x}_{0}$, then the closed-loop
system with $\Delta(s)$ is said to have the $\mathcal{L_{\text{\ensuremath{2}}}}$
gain margin $\gamma_{\text{max}}$. Given any $\Delta(s)=\text{diag}\left(\text{e}^{-s\tau_{1}}-1,\text{e}^{-s\tau_{2}}-1,\ldots,\text{e}^{-s\tau_{m}}-1\right)$,
$0\leq\tau_{i}\leq\tau_{\text{max}}$, $i=1,2\cdots m$, if the state
of the closed-loop system composed by (\ref{eq:Sys0}), (\ref{eq:UncertainyIn})
and $\mathbf{u}(t)=\mathbf{C}(\mathbf{x}(t))$ satisfies $\mathbf{x}\in\mathcal{L_{\text{\ensuremath{2}}}^{\text{\ensuremath{n}}}}$
for any initial condition $\mathbf{x}_{0}$, then the closed-loop
system with $\Delta(s)$ is said to have the $\mathcal{L_{\text{\ensuremath{2}}}}$
time-delay margin $\tau_{\text{max}}$.

\textbf{Objective 2} The other objective is to determine the $\mathcal{L_{\text{\ensuremath{2}}}}$
gain margin and $\mathcal{L_{\text{\ensuremath{2}}}}$ time-delay
margin for the resulting closed-loop nonlinear system.

\section{State-Compensation-Linearization-Based Stabilizing Control}

\label{Sec:Control}

The state-compensation-linearization-based\textit{\emph{ }}(or say
\textit{\emph{additive-state-decomposition-based}})\textit{\emph{
stabilizing control framework}} proposed in \cite{quan2024state}
can be used to propose a controller $\mathbf{u}(t)=\mathbf{C}(\mathbf{x}(t))$
to make $\mathbf{x}\in\mathcal{L_{\text{\ensuremath{2}}}^{\text{\ensuremath{n}}}}$
under any initial condition $\mathbf{x}_{0}$ when $\Delta(s)=\mathbf{0}$.
In the following, for convenience, we will omit the variables $t,s$
except when necessary.

\subsection{Additive State Decomposition}

Consider system (\ref{eq:Sys0}) as the original system. First, the
primary system is chosen as follows: 
\begin{equation}
\mathbf{\dot{x}}_{\text{p}}=\mathbf{A}\mathbf{x}_{\text{p}}+\mathbf{B}\mathbf{\boldsymbol{\mu}}_{\text{p}},\mathbf{x}_{\text{p}}(0)=\mathbf{x}_{0}\label{eq:Sys_Pri}
\end{equation}
where 
\begin{equation}
\mathbf{\boldsymbol{\mu}}_{\text{p}}=(\mathbf{I}_{{m}}+\Delta)(\mathbf{u}_{\text{p}}+\mathbf{u}_{\text{s}})-\mathbf{u}_{\text{s}}\label{eq:MuP}
\end{equation}
and $\mathbf{u}_{\text{s}}\triangleq\mathbf{u}-\mathbf{u}_{\text{p}}\in\mathbb{R}^{{m}}$
is the control for the secondary system in the following. Then, the
secondary system is determined by subtracting the primary system (\ref{eq:Sys_Pri})
from the original system (\ref{eq:Sys0}), and it follows that 
\begin{equation}
\mathbf{\dot{x}}_{\text{s}}=\mathbf{A\mathbf{x}}_{\text{s}}+\mathbf{f}(\mathbf{x}_{\text{p}}+\mathbf{x}_{\text{s}})+\mathbf{B}\mathbf{u}_{\text{s}},\mathbf{x}_{\text{s}}(0)=\mathbf{0}.\label{eq:Sys_Sec}
\end{equation}
If $\mathbf{x}_{\text{p}}=\mathbf{0}$, then $(\mathbf{x}_{\text{s}},\mathbf{u}_{\text{s}})=\mathbf{0}$
is an equilibrium point of (\ref{eq:Sys_Sec}).

An observer is proposed to estimate $\mathbf{x}_{\text{p}}$ and $\mathbf{x}_{\text{s}}$
in (\ref{eq:Sys_Pri}), (\ref{eq:Sys_Sec}) as follows: 
\begin{align}
\dot{\mathbf{\mathbf{\hat{\mathbf{x}}}}}_{\text{s}} & =\mathbf{A}\hat{\mathbf{x}}_{\text{s}}+\mathbf{f}(\mathbf{x})+\mathbf{B}\mathbf{\mathbf{u}_{\text{s}}},\hat{\mathbf{x}}_{\text{s}}(0)=\mathbf{0}\label{eq:Observer_Eq1}\\
\hat{\mathbf{x}}_{\text{p}} & =\mathbf{x}-\mathbf{\hat{x}}_{\text{s}}.\label{eq:Observer_Eq2}
\end{align}
Then $\hat{\mathbf{x}}_{\text{p}}=\mathbf{x}_{\text{p}}$ and $\mathbf{\hat{x}}_{\text{s}}=\mathbf{x}_{\text{s}}$.
The proof can refer to \cite{quan2024state}.

\subsection{Stabilizing Control}

In the control design, we first ignore the unmodeled dynamics $\Delta$
in (\ref{eq:Sys0}), namely $\Delta=\mathbf{0}$ (\textit{Objective
1}). However, it will appear in the stability margin analysis (\textit{Objective
2}). When $\Delta=\mathbf{0}$, in the primary system (\ref{eq:Sys_Pri}),
we have

\[
\mathbf{\boldsymbol{\mu}}_{\text{p}}=\mathbf{u}_{\text{p}}.
\]

\begin{itemize}
\item \textbf{Problem 1} Consider the primary system, and the dynamic of
which is described by (\ref{eq:Sys_Pri}) with $\Delta=\mathbf{0}$.
Design the primary controller as
\begin{equation}
\mathbf{u}_{\text{p}}=\mathcal{L}^{-1}(\mathbf{H}(s)\mathbf{K}\mathbf{x_{\text{p}}}(s))\label{eq:Controller_Prob1-1}
\end{equation}
such that $\mathbf{x_{\text{p}}}\in\mathcal{L_{\text{\ensuremath{2}}}^{\text{\ensuremath{n}}}}$,
where $\text{\textbf{H}(s)}$ is a stable transfer function matrix,
$\mathcal{L}^{-1}$ denotes the inverse Laplace transformation, and
state feedback matrix $\mathbf{K\in\mathbb{R}^{\text{\ensuremath{m\times n}}}}$. 
\item \textbf{Problem 2} Consider the secondary system, and the dynamic
of which is described by (\ref{eq:Sys_Sec}). Design the secondary
controller as
\begin{equation}
\mathbf{u}_{\text{s}}=\mathbf{L}(\mathbf{x_{\text{p}}},\mathbf{x_{\text{s}}})\label{eq:Controller_Prob2-1}
\end{equation}
such that $\mathbf{x_{\text{s}}}\in\mathcal{L_{\text{\ensuremath{2}}}^{\text{\ensuremath{n}}}}$,
where $\mathbf{L}(\cdot,\cdot)$ is a nonlinear function. 
\end{itemize}
With the solutions to the two problems in hand, we can state

\textbf{Theorem 1} For system (\ref{eq:Sys0}) with $\Delta=\mathbf{0}$
under \emph{Assumptions 1-2}, suppose (i) \emph{Problems 1-2} are
solved; (ii) the controller for system (\ref{eq:Sys0}) with $\Delta=\mathbf{0}$
is designed as 
\begin{align}
\mathbf{\dot{\mathbf{\mathbf{\hat{x}}}}}_{\text{s}} & =\mathbf{A}\hat{\mathbf{x}}_{\text{s}}+\mathbf{f}(\mathbf{x})+\mathbf{B}\mathbf{\mathbf{u}_{\text{s}}},\hat{\mathbf{x}}_{\text{s}}(0)=\mathbf{0}\nonumber \\
\hat{\mathbf{x}}_{\text{p}} & =\mathbf{x}-\mathbf{\hat{x}}_{\text{s}}\nonumber \\
\mathbf{\boldsymbol{\mu}} & =\mathbf{C}\left(\mathbf{x}\right)=\mathcal{L}^{-1}(\mathbf{H}(s)\mathbf{K}\hat{\mathbf{x}}_{\text{p}}(s))+\mathbf{L}(\hat{\mathbf{x}}_{\text{p}},\mathbf{\hat{x}}_{\text{s}}).\label{eq:Controller}
\end{align}
Then, the state of system (\ref{eq:Sys0}) with $\Delta=\mathbf{0}$
satisfies $\mathbf{x}\in\mathcal{L_{\text{\ensuremath{2}}}^{\text{\ensuremath{n}}}}$.

\begin{figure}[tbh]
\begin{centering}
\includegraphics[width=0.5\textwidth]{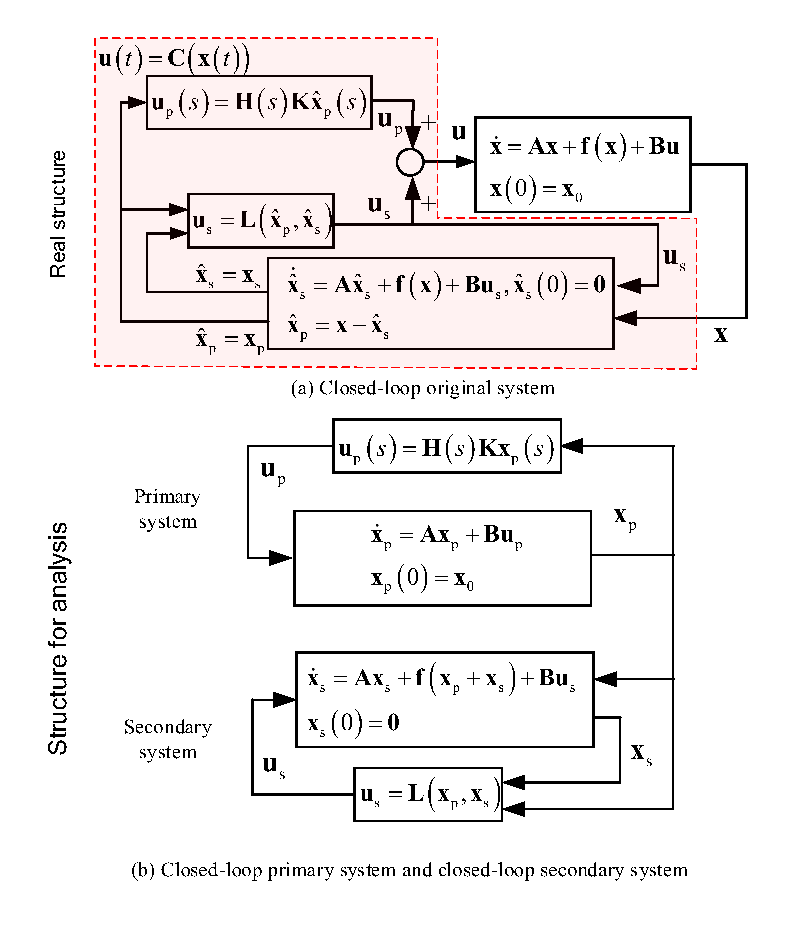}
\par\end{centering}
\caption{When unmodeled dynamics $\Delta=\mathbf{0}$, Fig.\,\ref{Fig_closed-loop}(a)
shows the real closed-loop system corresponding to the original system,
which can be decomposed into the two closed-loop subsystems in Fig.\,\ref{Fig_closed-loop}(b)
for analysis.}
\label{Fig_closed-loop}
\end{figure}

The controllers for the primary system and the secondary system are
shown in Fig.\,\ref{Fig_closed-loop}(b) for analysis. As shown,
when $\Delta=\mathbf{0}$, the primary system is independent of the
secondary system. However, the secondary system depends on the state
of the primary system. In practice, the two controllers are combined
to control the original system as shown in Fig.\,\ref{Fig_closed-loop}(a).

So far, we have solved \emph{Objective 1} by \emph{Theorem 1} and
derived the controller $\mathbf{C}\left(\mathbf{x}\right)$ as in
(\ref{eq:Controller}). However, it is not sufficient just know whether
a system is stable or unstable. If a system is just barely stable,
then a small perturbation in a system parameter could push the system
over the edge. Thus, one often wants to design systems with some robust
stability margin. In the following, the $\mathcal{L_{\text{\ensuremath{2}}}}$
gain margin and $\mathcal{L_{\text{\ensuremath{2}}}}$ time-delay
margin will be used to measure how stable a system is, and how to
determine them is further concerned (\emph{Objective 2}).

\section{State-Compensation-Linearization-Based Stability Margin Analysis}

\label{sec:SM}

\subsection{Stability Margin Analysis}

\begin{figure}[tbh]
\begin{centering}
\includegraphics[width=0.5\textwidth]{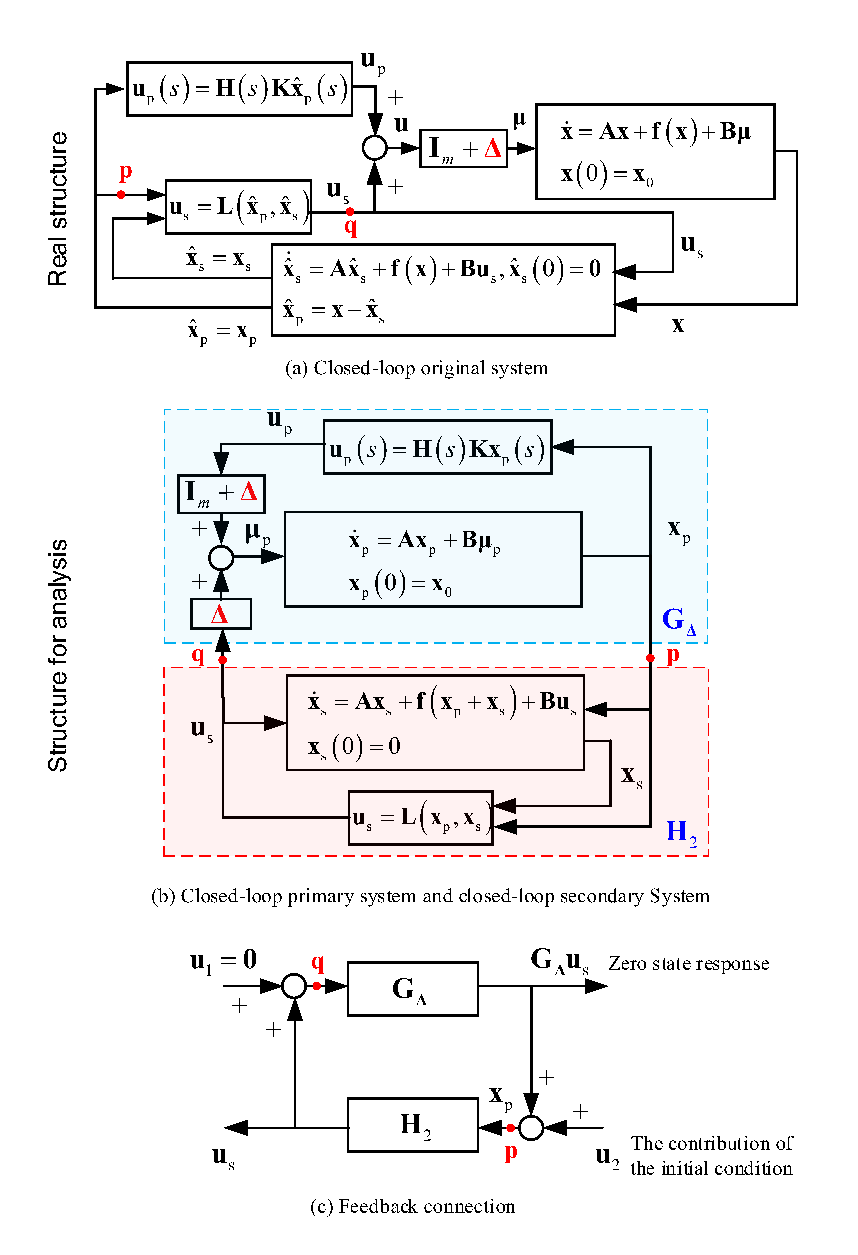}
\par\end{centering}
\caption{In the presence of unmodeled dynamics $\Delta\protect\ne\mathbf{0}$,
Fig.\,\ref{Fig_closed-loopU}(a) shows the real closed-loop system
corresponding to the original system, which can be transformed into
Fig.\,\ref{Fig_closed-loopU}(b) for analysis. Furthermore, Fig.\,\ref{Fig_closed-loopU}(c)
is an abstract feedback connection of Fig.\,\ref{Fig_closed-loopU}(b).}
\label{Fig_closed-loopU}
\end{figure}

From the definition of $\mathbf{\boldsymbol{\mu}}_{\text{p}}$ in
(\ref{eq:MuP}), we have

\[
\mathbf{\boldsymbol{\mu}}_{\text{p}}=(\mathbf{I}_{{m}}+\Delta)\mathbf{u}_{\text{p}}+\Delta\mathbf{u}_{\text{s}}.
\]
In the presence of $\Delta$, Fig.\,\ref{Fig_closed-loop} is modified
to Fig.\,\ref{Fig_closed-loopU}. Because of $\Delta$, the primary
system and the secondary system are interconnected, namely the primary
system is dependent on the secondary system and the secondary system
is, in turn, dependent on the primary system as shown in Fig.\,\ref{Fig_closed-loopU}(b).
What is more, the interconnection of the primary system and the secondary
system can be converted to the form in Fig.\,\ref{Fig_closed-loopU}(c),
where $\mathbf{u}_{\text{1}}=\mathbf{0}$ and $\mathbf{u}_{\text{2}}$
is the contribution of the initial condition $\mathbf{x_{\text{0}}}$.
Since $\mathbf{u}_{\text{2}}$ can be taken as an impulse at time
$t=0$, which belongs to $\mathcal{L_{\text{\ensuremath{2}}}^{\text{\ensuremath{n}}}}$,
then $\mathbf{u}=\left[\mathbf{u_{\text{1}}^{\text{T}}}\:\mathbf{u_{\text{2}}^{\text{T}}}\right]^{\text{T}}\in\mathcal{L}_{\text{\ensuremath{2}}}^{\text{\ensuremath{n}}}$.

By exploiting the interconnection structure and the small-gain theorem,
the following theorem is drawn to prove the robust stability.

\textbf{Theorem 2} For system (\ref{eq:Sys0}) with $\Delta$ under
\emph{Assumptions 1-2}, suppose (i) \emph{Problems 1-2} are solved;
(ii) the controller for system (\ref{eq:Sys0}) with $\Delta=\mathbf{0}$
is designed as in (\ref{eq:Controller}); (iii) there exist $k_{l}>0$
such that $\left\Vert \mathbf{u}_{\text{s}}\right\Vert _{\mathcal{L_{\text{\ensuremath{2}}}}}\leq k_{l}\left\Vert \mathbf{x_{\text{p}}}\right\Vert _{\mathcal{L_{\text{\ensuremath{2}}}}}+\beta$,
where , $\beta\in\mathcal{L}_{\text{\ensuremath{2}}}^{\text{\ensuremath{n}}}$
is a constant that depends on the initial state; (iv) $\mathbf{G}_{\Delta}$
is finite-gain $\mathcal{L}_{\text{\ensuremath{2}}}$ stable, namely
$\mathbf{G}_{\Delta}:\mathcal{L}_{\text{\ensuremath{2}}}^{\text{\ensuremath{m}}}\rightarrow\mathcal{L}_{\text{\ensuremath{2}}}^{\text{\ensuremath{n}}}$
and

\begin{equation}
k_{l}\left\Vert \mathbf{G}_{\Delta}\right\Vert <1\label{eq:Smallg4Osys}
\end{equation}
where

\[
\mathbf{G}_{\Delta}(s)=(s\mathbf{I}_{n}-\mathbf{A}-(\mathbf{I}_{m}+\Delta(s))\mathbf{H}(s)\mathbf{B}\mathbf{K})^{-1}\mathbf{B}\Delta(s).
\]
Then, with controller (\ref{eq:Controller}), the state of system
(\ref{eq:Sys0}) satisfies $\mathbf{x}\in\mathcal{L}_{\text{\ensuremath{2}}}^{\text{\ensuremath{n}}}$.

\emph{Proof}. On the one hand,\emph{ Problem 2} for the secondary
system is solved and 

\begin{equation}
\left\Vert \mathbf{u}_{\text{s}}\right\Vert _{\mathcal{L_{\text{\ensuremath{2}}}}}\leq k_{l}\left\Vert \mathbf{x_{\text{p}}}\right\Vert _{\mathcal{L_{\text{\ensuremath{2}}}}}+\beta.\label{eq:E1}
\end{equation}
On the other hand, based on (\ref{eq:Controller_Prob1-1}), (\ref{eq:Sys_Pri}),
and (\ref{eq:MuP}), we have

\[
\mathbf{x_{\text{p}}}=(s\mathbf{I}_{n}-\mathbf{A})^{-1}\mathbf{B}(\Delta\mathbf{u}_{\text{s}}+(\mathbf{I}_{m}+\Delta(s))\mathbf{H}(s)\mathbf{K}\mathbf{x_{\text{p}}})+(s\mathbf{I}_{n}-\mathbf{A})^{-1}\mathbf{x_{\text{0}}}.
\]
Then, as shown in Fig.\,\ref{Fig_closed-loopU}(c), we have

\begin{equation}
\mathbf{x_{\text{p}}}=\mathbf{G}_{\Delta}\mathbf{u}_{\text{s}}+\mathbf{u}_{\text{2}}\label{eq:SysGp}
\end{equation}
where $\mathbf{u}_{\text{2}}=(s\mathbf{I}_{n}-\mathbf{A}-(\mathbf{I}_{m}+\Delta(s))\mathbf{H}(s)\mathbf{B}\mathbf{K})^{-1}\mathbf{x_{\text{0}}}$.
Since $\mathbf{G}_{\Delta}:\mathcal{L}_{\text{\ensuremath{2}}}^{\text{\ensuremath{m}}}\rightarrow\mathcal{L}_{\text{\ensuremath{2}}}^{\text{\ensuremath{n}}},\mathbf{u}_{\text{2}}\mathcal{\in L}_{\text{\ensuremath{2}}}^{\text{\ensuremath{n}}}$.
Consequently,

\begin{equation}
\left\Vert \mathbf{x_{\text{p}}}\right\Vert _{\mathcal{L_{\text{\ensuremath{2}}}}}\le\left\Vert \mathbf{G}_{\Delta}\right\Vert \left\Vert \mathbf{u}_{\text{s}}\right\Vert _{\mathcal{L_{\text{\ensuremath{2}}}}}+\left\Vert \mathbf{u}_{\text{2}}\right\Vert _{\mathcal{L_{\text{\ensuremath{2}}}}}.\label{eq:E2}
\end{equation}
Based on (\ref{eq:E1}) and (\ref{eq:E2}), if the condition (\ref{eq:Smallg4Osys})
is satisfied, then the interconnection shown in Fig.\,\ref{Fig_closed-loopU}(c)
is finite-gain $\mathcal{L}_{\text{\ensuremath{2}}}$ stable with
the input $\mathbf{u}=\left[\mathbf{u_{\text{1}}^{\text{T}}}\:\mathbf{u_{\text{2}}^{\text{T}}}\right]^{\text{T}}\in\mathcal{L}_{\text{\ensuremath{2}}}^{n+m}$
(see the small-gain theorem). Since $\mathbf{u}\in\mathcal{L}_{\text{\ensuremath{2}}}^{\text{\ensuremath{n+m}}},\mathbf{u}_{\text{s}}\mathcal{\in L}_{\text{\ensuremath{2}}}^{\text{\ensuremath{m}}}$
and $\mathbf{x_{\text{p}}}\in\mathcal{L}_{\text{\ensuremath{2}}}^{\text{\ensuremath{n}}}$.
In addition, $\mathbf{x_{\text{s}}}\in\mathcal{L}_{\text{\ensuremath{2}}}^{\text{\ensuremath{n}}}$
can be guaranteed by \emph{Problem 2}. Therefore, according to additive
state decomposition, $\mathbf{x}\in\mathcal{L}_{\text{\ensuremath{2}}}^{\text{\ensuremath{n}}}$.
$\Square$

With \emph{Theorem 2} in hand, we further have the following theorem
for stability margin.

\textbf{Theorem 3} (Stability Margin). For system (\ref{eq:Sys0})
with $\Delta$ under \emph{Assumptions 1-2}, suppose (i) \emph{Problems
1-2} are solved; (ii) the controller for system (\ref{eq:Sys0}) with
$\Delta=\mathbf{0}$ is designed as in (\ref{eq:Controller}). Then

(i) $\mathcal{L}_{\text{\ensuremath{2}}}$ \emph{gain margin} of the
closed-loop system, $\gamma_{\text{max}}>0$, is such that for any
$\left|\gamma_{i}\right|\leq\gamma_{\text{max}}$, we have

\begin{equation}
k_{l}\left\Vert \mathbf{G}_{\gamma}\right\Vert <1\label{eq:Gain}
\end{equation}
where $\mathbf{G}_{\gamma}(s)=(s\mathbf{I}_{n}-\mathbf{A}-(\mathbf{I}_{m}+\mathbf{E}_{1})\mathbf{H}(s)\mathbf{B}\mathbf{K})^{-1}\mathbf{B}\mathbf{E}_{1}$,
$\mathbf{E}_{1}=\text{diag}\left(\gamma_{1},\gamma_{2},...,\gamma_{m}\right)$.

(ii) $\mathcal{L}_{\text{\ensuremath{2}}}$ \emph{time-delay margin}
of the closed-loop system, $\tau_{\text{max}}>0$, is such that for
any $0\leq\tau_{i}\leq\tau_{\text{max}}$, we have

\begin{equation}
k_{l}\left\Vert \mathbf{G}_{\tau}\right\Vert <1\label{eq:Delay}
\end{equation}
where $\mathbf{G}_{\tau}(s)=(s\mathbf{I}_{n}-\mathbf{A}-(\mathbf{I}_{m}+\text{\ensuremath{\mathbf{E}_{2}}})\mathbf{H}(s)\mathbf{B}\mathbf{K})^{-1}\mathbf{B}\mathbf{E}_{2}$,
$\mathbf{E}_{2}=\text{diag\ensuremath{\left(\text{e}^{-s\tau_{1}}-1,...,\text{e}^{-s\tau_{m}}-1\right)}})$.

\emph{Proof}. For the $\mathcal{L}_{\text{\ensuremath{2}}}$ gain
margin, let $\Delta=\textrm{diag}\left(\gamma_{1},\gamma_{2},...,\gamma_{m}\right)$
in (\ref{eq:UncertainyIn}). Then, according to \emph{Theorem 2} and
\emph{Definition 1}, we can prove (i). Similarly, for the $\mathcal{L}_{\text{\ensuremath{2}}}$
time-delay margin, let $\Delta=\text{diag\ensuremath{\left(\text{e}^{-s\tau_{1}}-1,...,\text{e}^{-s\tau_{m}}-1\right)}}$
in (\ref{eq:UncertainyIn}). Then, according to \emph{Theorem 2} and
\emph{Definition 1}, we can prove (ii).$\Square$

\subsection{Stability Margin Measurement}

\label{Sec:SM Measurement}

The $\mathcal{L}_{\text{\ensuremath{2}}}$ gain margin and $\mathcal{L}_{\text{\ensuremath{2}}}$
time-delay margin can be calculated theoretically according to the
given parameters and \emph{Theorem 3.} In the following, we will present
a frequency-sweep method to acquire them. This offers a way to get
the stability margin through real experiments rather than theoretical
calculation, which approaches reality more.

\subsubsection{Stability Margin for the Primary System}

\label{Sec:SM PSys}

We first should find $\gamma_{\text{max,1}}$ and $\tau_{\text{max,1}}$
to make $\mathbf{G}_{\Delta}:\mathcal{L}_{\text{\ensuremath{2}}}^{\text{\ensuremath{m}}}\rightarrow\mathcal{L}_{\text{\ensuremath{2}}}^{\text{\ensuremath{n}}}$,
namely the stability condition for the primary system (the upper half
part of the system shown in Fig. 4(b)). Because the primary system
is linear, the proposed stability margin analysis method in \cite{https://doi.org/10.1002/rnc.7413}can
be used. The stability margin is obtained by measuring the upper half
part of the system shown in Fig. 4(b). For SISO systems, the classic
stability margin is considered; for MIMO systems, the $\mathcal{L_{\text{\ensuremath{2}}}}$
stability margin is considered.

\subsubsection{Stability Margin for the Whole System}

\label{Sec:SM OSys}

Based on the results above, we should further find $\gamma_{\text{max,2}}$
and $\tau_{\text{max,2}}$ to make $k_{l}\left\Vert \mathbf{G}_{\Delta}\right\Vert <1$,
namely the stability condition for the whole system.

For the $\mathcal{L}_{\text{\ensuremath{2}}}$ gain margin, according
to $\left(\ref{eq:Gain}\right)$, if

\[
\gamma_{\text{max,2}}<\left\Vert \mathbf{G}_{0}\mathbf{B}\right\Vert ^{-1}\left(k_{l}\right)^{-1}
\]
then $\gamma\in\left[-\gamma_{\text{max,2}},\:\gamma_{\text{max,2}}\right]$
can make $k_{l}\left\Vert \mathbf{G}_{\gamma}\right\Vert <1$. Here

\begin{equation}
\gamma_{\text{max,2}}=\left(1-\varepsilon_{3}\right)\left\Vert \mathbf{G}_{0}\mathbf{B}\right\Vert ^{-1}\left(k_{l}\right)^{-1}\label{eq:Gamma2}
\end{equation}
where $0<\varepsilon_{3}<1$ is often chosen sufficiently small. For
the $\mathcal{L}_{\text{\ensuremath{2}}}$ time-delay margin, we have

\[
\tau_{\text{max,2}}<\left\Vert s\mathbf{G}_{0}\mathbf{B}\right\Vert ^{-1}\left(k_{l}\right)^{-1}
\]
then $\tau\in\left[0,\:\tau_{\text{max,2}}\right]$ can make $k_{l}\left\Vert \mathbf{G}_{\tau}\right\Vert <1$.
Here

\begin{equation}
\tau_{\text{max,2}}=\left(1-\varepsilon_{4}\right)\left\Vert s\mathbf{G}_{0}\mathbf{B}\right\Vert ^{-1}\left(k_{l}\right)^{-1}\label{eq:Tao2}
\end{equation}
where $0<\varepsilon_{4}<1$ is often chosen sufficiently small.

\begin{figure}
\begin{centering}
\includegraphics[width=0.5\textwidth]{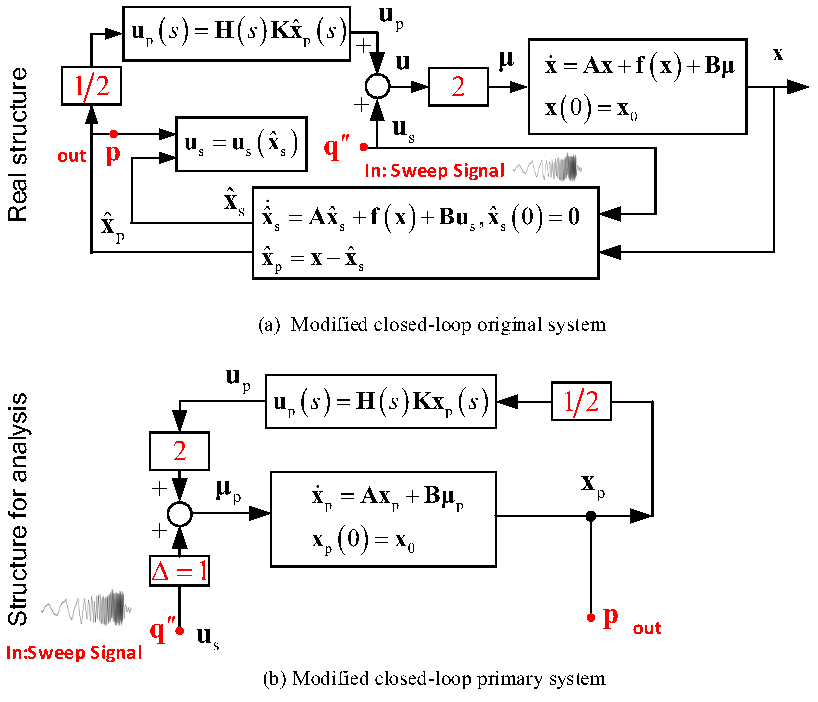}
\par\end{centering}
\caption{Establishing $\mathbf{G}_{0}\mathbf{B}$. From the input-output perspective,
the open-loop systems from $\mathbf{q}^{\prime\prime}$ to $\mathbf{p}$
in Fig.\,\ref{Fig_SweepG4OSys}(a) and Fig.\,\ref{Fig_SweepG4OSys}(b)
are equivalent.}
\label{Fig_SweepG4OSys}
\end{figure}

In the following, we aim to get $\left\Vert \mathbf{G}_{0}\mathbf{B}\right\Vert $
and $\left\Vert s\mathbf{G}_{0}\mathbf{B}\right\Vert $ by the frequency-sweep
method. To obtain $\mathbf{G}_{0}\mathbf{B}$, we should modify Fig.\,\ref{Fig_closed-loopU}
to be Fig.\,\ref{Fig_SweepG4OSys}. In Fig.\,\ref{Fig_SweepG4OSys},
the open loops from $\mathbf{q}^{\prime\prime}$ to $\mathbf{p}$
for Fig.\,\ref{Fig_SweepG4OSys}(a) (real structure) and Fig.\,\ref{Fig_SweepG4OSys}(b)
(structure for analysis) are the same. In Fig.\,\ref{Fig_SweepG4OSys}(b),
the transfer function from $\mathbf{q}^{\prime\prime}$ to $\mathbf{p}$
is $\mathbf{G}_{0}\mathbf{B}$. The frequency-sweep method is only
applicable to the open loop from $\mathbf{q}^{\prime\prime}$ to $\mathbf{p}$
in Fig.\,\ref{Fig_SweepG4OSys}(a), because the open loop in Fig.\,\ref{Fig_SweepG4OSys}(b)
is conceived only for controller design and analysis. The procedure
to get $\mathbf{G}_{0}\mathbf{B}$ is as follows: 
\begin{itemize}
\item Recalling the system with $\Delta$ as shown in Fig.\,\ref{Fig_closed-loopU}(b),
it can be seen that the terms $\Delta$ and $\mathbf{I}_{m}+\Delta$
will make an obstacle to obtaining $\mathbf{G}_{0}\mathbf{B}$ from
$\mathbf{q}$ to $\mathbf{p}$. 
\item Let $\Delta=\mathbf{I}_{m}$ and insert a gain matrix $1/2\mathbf{I}_{m}$
into the closed-loop primary system in Fig.\,\ref{Fig_closed-loopU}(b).
Then, in practice, $\mathbf{G}_{0}\mathbf{B}$ is obtained from $\mathbf{q}^{\prime\prime}$
to $\mathbf{p}$ in the way shown in Fig.\,\ref{Fig_SweepG4OSys}(a). 
\end{itemize}
It should be noticed that these additional terms are only used to
obtain $\mathbf{G}_{0}\mathbf{B}$, where $\Delta=\mathbf{I}_{m}$
is a simple way to establish $\mathbf{G}_{0}\mathbf{B}$.

As shown in Fig.\,\ref{Fig_SweepG4OSys}(b), the output responses
are collected at the point $\mathbf{p}$ by feeding frequency-sweep
inputs (sweeping the frequency of input signals from 0 to a value
just beyond the frequency range of the system) to the system at the
point $\mathbf{q}^{\prime\prime}$ . Based on the input-output data,
one can get $\left\Vert \mathbf{G}_{0}\mathbf{B}\right\Vert $ and
$\left\Vert s\mathbf{G}_{0}\mathbf{B}\right\Vert $. After getting
$\left\Vert \mathbf{G}_{0}\mathbf{B}\right\Vert $ and $\left\Vert s\mathbf{G}_{0}\mathbf{B}\right\Vert $
by the frequency-sweep method, $\gamma_{\text{max,2}}$ and $\tau_{\text{max,2}}$
can be obtained by (\ref{eq:Gamma2}), (\ref{eq:Tao2}), namely the
stability condition for the whole system is determined.

\subsubsection{Stability Margin}

Finally, with $\gamma_{\text{max,1}}$, $\tau_{\text{max,1}}$, $\gamma_{\text{max,2}}$
and $\tau_{\text{max,2}}$ available, the $\mathcal{L}_{\text{\ensuremath{2}}}$
gain margin is

\[
\gamma_{\text{max}}\thickapprox\min\left(\gamma_{\text{max,1}},\gamma_{\text{max,2}}\right)
\]
and the $\mathcal{L}_{\text{\ensuremath{2}}}$ time-delay margin is

\[
\tau_{\text{max}}\thickapprox\min\left(\tau_{\text{max,1}},\tau_{\text{max,2}}\right).
\]

\subsection{Whole Design and Analysis Procedure}

\begin{figure*}
\centering\includegraphics[width=0.7\textwidth]{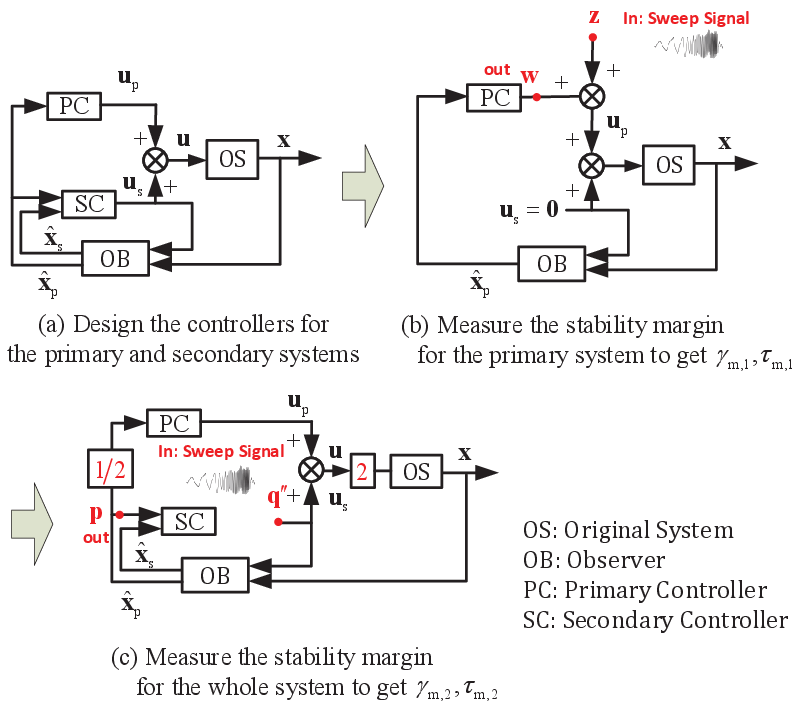}\caption{Procedure to design the controller and measure the stability margin.}
\label{Fig_Procedure}
\end{figure*}

The complete procedure to design the controller and measure the stability
margin is summarized in Fig.\,\ref{Fig_Procedure}. The procedure
consists of three steps, in which Step (a) belongs to the state-compensation-linearization-based
control design (corresponding to Section \ref{Sec:Control}); Step
(b) works for the stability margin measurement for the primary system
(corresponding to Section \ref{Sec:SM PSys}); Step (c) aims to measure
the stability margin for the whole system (corresponding to Section
\ref{Sec:SM OSys}). This approach has the advantage of getting stability
margin through data by real experiments, which takes both model uncertainties
and external disturbances into account, and thus approaches to practice
more. 

\section{Simulation Studies}

\label{sec:Simulations}

In order to show the effectiveness of the state-compensation-linearization-based
stability margin analysis method proposed above, three illustrative
examples are considered in the simulation.

\subsection{Example 1: An SISO nonlinear system}

This example aims to demonstrate the complete procedure to design
the controller and measure the stability margin for an SISO nonlinear
system by using the proposed method, and to show the difference between
the defined stability margin and the classic frequency-domain stability
margin. Consider the following second-order nonlinear system 
\begin{align}
\dot{x}_{1} & =x_{2}\nonumber \\
\dot{x}_{2} & =-2x_{1}-3x_{2}+\frac{x_{2}^{2}}{1+ax_{2}^{2}}+\mu\label{Sys1}
\end{align}
with 
\[
\mu=\left(1+\Delta\right)u
\]
where $a>0$. The system can be transformed into the form in (\ref{eq:Sys0})
with 
\[
\mathbf{A=}\left[\begin{array}{cc}
0 & 1\\
-2 & -3
\end{array}\right],\mathbf{f}\left(\mathbf{x}\right)=\left[\begin{array}{c}
0\\
\frac{x_{2}^{2}}{1+ax_{2}^{2}}
\end{array}\right],\mathbf{b}=\left[\begin{array}{c}
0\\
1
\end{array}\right].
\]
It is obvious that $\mathbf{A}$ is stable.

\subsubsection{State-compensation-linearization-based control (SCLC)}

In this part, the controller design and stability margin analysis
will be carried out by following the procedure illustrated in Fig.\ \ref{Fig_Procedure}.

$\bullet$ Step (a). Design controllers for the primary system and
the secondary system. For the controller (\ref{eq:Controller_Prob1-1}),
the LQR method is employed to calculate the feedback matrix $\mathbf{k}$.\textbf{
}Besides, let $h\left(s\right)=1$. Then, the primary controller follows
\[
u_{\text{p}}=\mathbf{k}^{\text{T}}\mathbf{\hat{x}}_{\text{p}}.
\]
For the controller (\ref{eq:Controller_Prob2-1}), the backstepping
control is adopted, and a secondary controller is designed as 
\begin{equation}
u_{\text{s}}=\left(3-c_{1}-c_{2}\right)\hat{x}_{\text{s},2}+\left(1-c_{1}c_{2}\right)\hat{x}_{\text{s},1}-\frac{(\hat{x}_{\text{p},2}+\hat{x}_{\text{s},2})^{2}}{1+a(\hat{x}_{\text{p},2}+\hat{x}_{\text{s},2})^{2}}\label{US}
\end{equation}
where $c_{1},c_{2}>0$ are parameters to be specified later. The concrete
design procedure for backstepping control is omitted here for it is
straightforward by routine steps.

$\bullet$ Step (b). Measure the stability margin for the primary
system to get $\gamma_{\text{max},1}$ and $\tau_{\text{max},1}$.
Because the primary system is an SISO system, the classic stability
margin can be considered. The frequency-sweep method is utilized here.
Based on the input-output data, the Bode plot of the open-loop system
is obtained. Then, the GM $G_{\text{m},1}$, the PM $P_{\text{m},1}$,
and the gain and phase crossover frequencies $\omega_{\text{cg},1},\omega_{\text{cp},1}$
are obtained relying on the Bode plot. Finally, $\gamma_{\text{max},1}$
and $\tau_{\text{max},1}$ are determined.

$\bullet$ Step (c). Measure the stability margin for the whole system
to get $\gamma_{\text{max},2}$ and $\tau_{\text{max},2}$. First,
$k_{l}$ is calculated based on (\ref{US}). Then, based on Section
\ref{Sec:SM OSys}, the frequency-sweep test is carried out by applying
frequency-sweep input signals to the system at the point $\mathbf{q}^{\prime\prime}$
and collecting output responses at the point $\mathbf{p}$, as shown
in Fig.\ \ref{Fig_Procedure}(c). By the corresponding collected
input-output data, $\left\Vert \mathbf{G}_{0}\mathbf{b}\right\Vert $
and $\left\Vert s\mathbf{G}_{0}\mathbf{b}\right\Vert $ are gotten,
and then $\gamma_{\text{max},2}$ and $\tau_{\text{max},2}$ are obtained
by (\ref{eq:Gamma2}) and (\ref{eq:Tao2}).

\subsubsection{JLC for comparison}

A commonly-used way is to apply linear control techniques at discrete
operating points within the operating envelope, and then the classic
frequency-domain stability margin is used to evaluate the designed
controllers at these discrete points. For stabilizing control, the
operating point is the equilibrium point, namely, the origin herein.
By linearizing the system (\ref{Sys1}) at the origin $\mathbf{x}=\mathbf{0}$,
one has 
\begin{align}
\dot{x}_{1} & =x_{2}\nonumber \\
\dot{x}_{2} & =-2x_{1}-3x_{2}+\mu.\label{LSys}
\end{align}
Then, similar to the controller (\ref{eq:Controller_Prob1-1}), an
LQR-based stabilizing controller for (\ref{LSys}) is designed as
\begin{equation}
u=\mathbf{k}^{\text{T}}\mathbf{x.}\label{DLBC1}
\end{equation}
where the parameter $\mathbf{k}$ is chosen the same as the state-compensation-linearization-based
control mentioned above.

The stability margin analysis for the closed-loop system consisting
of (\ref{LSys}) and (\ref{DLBC1}) can be done by a classic frequency-domain
method, which is the same as that of the primary system in the state-compensation-linearization-based
control.

\subsubsection{Simulation results}

\begin{figure}[ptbh]
\begin{centering}
\includegraphics[width=0.45\textwidth]{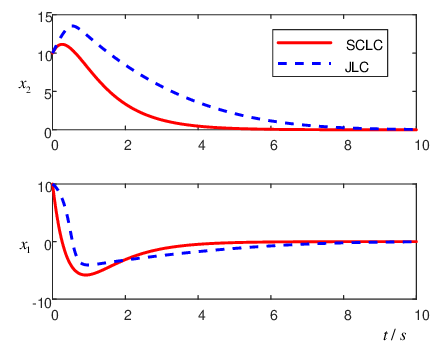}
\par\end{centering}
\caption{State response (Example 2).}
\label{SC_Effect}
\end{figure}

For a fair comparison, the parameters in LQR are selected as the same
values for SCLC and JLC. In the following simulations, $\mathbf{Q}=$diag$\left(1\text{, }1\right),R=1$
are selected, resulting in $\mathbf{k=-}\left[0.236\text{ }0.236\right]^{\text{T}}$.
Besides, let $a=0.01,\mathbf{x}_{0}=[10$ $10]^{\text{T}},c_{1},c_{2}=20$.
The corresponding stabilizing control performance is compared in Fig.\ \ref{SC_Effect}.
It can be observed that SCLC outperforms JLC. 

\begin{figure}[ptbh]
\begin{centering}
\includegraphics[width=0.5\textwidth]{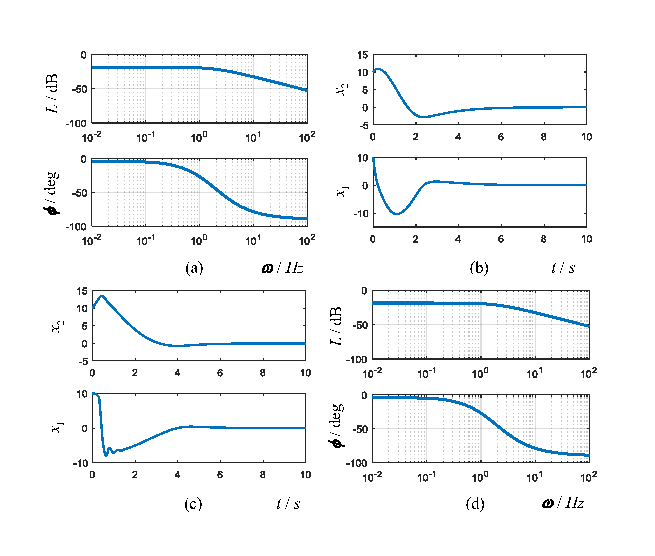}
\par\end{centering}
\caption{Stability margin analysis results (Example 2): (a) Bode plot for the
primary system in SCLC; (b) Validation test results of $\gamma_{\text{max}}$;
(c) Validation test results of $\tau_{\text{max}}$; (d) Bode plot
in JLC.}
\label{Ex1}
\end{figure}

In the following, the stability margin analysis is performed after
the stabilizing control has been done. For SCLC, the stability margin
for the primary system is studied firstly. The GM and the PM are obtained
as $G_{\text{m},1}=\infty$, $P_{\text{m},1}=\infty$ by the Bode
plot as represented in Fig.\ \ref{Ex1}(a), which comes from the
frequency-sweep test. Then, it is determined that $\gamma_{\text{max},1}=\infty$
and $\tau_{\text{max},1}=\infty$. Thus, the primary system is stable,
i.e., $\mathbf{G}_{\Delta}$: $\mathcal{L}_{2}^{1}\rightarrow\mathcal{L}_{2}^{n}$,
and has a large stability margin. Then, the stability margin for the
whole system is obtained in two ways: the theoretical computation
method and the frequency-sweep method. The theoretical computation
method is to find $\gamma_{\text{max},2}$ and $\tau_{\text{max},2}$
by gradually increasing them from $0$ until the inequality (\ref{eq:Smallg4Osys})
does not hold. On the other hand, the frequency-sweep method is to
get $\left\Vert \mathbf{G}_{0}\mathbf{b}\right\Vert $ and $\left\Vert s\mathbf{G}_{0}\mathbf{b}\right\Vert $
by the corresponding frequency-sweep data, and then $\gamma_{\text{max},2}$
and $\tau_{\text{max},2}$ can be obtained by equations (\ref{eq:Gamma2}),
(\ref{eq:Tao2}). Through theoretical computation, the stability margin
for the whole system is obtained as $\gamma_{\text{max},2}=0.428$
and $\tau_{\text{max},2}=0.22$, while through the frequency-sweep
method, it is obtained that $\gamma_{\text{max},2}=0.45$ and $\tau_{\text{max},2}=0.20$.
Note that the two groups of results show good agreement with each
other. Finally, $\gamma_{\text{max}}\approx\min\left(\gamma_{\text{max},1},\gamma_{\text{max},2}\right)=0.45$,
$\tau_{\text{max}}\approx\min\left(\tau_{\text{max},1},\tau_{\text{max},2}\right)=0.20$.

To verify the obtained $\gamma_{\text{max}}$, $\Delta=0.45$ is added
to the original system, and the corresponding state response is depicted
in Fig.\ \ref{Ex1}(b). To verify the obtained $\tau_{\text{max}}$,
$\Delta=e^{-0.2s}-1$ is added to the original system, and the corresponding
state response is depicted in Fig.\ \ref{Ex1}(c). It can be observed
that although the state responses all get worse compared to the results
in Fig.\ \ref{SC_Effect}, the system is still stable eventually.

For JLC, the current practice is to certify such systems by the GM
and PM at the origin. From Fig.\ \ref{Ex1}(d), the GM and PM are
$G_{\text{m}}=\infty$, $P_{\text{m}}=\infty,$ which are nearly the
same as the obtained stability margin results for the primary system
in SCLC. Actually, the method directly treats the nonlinear system
as a linear one, and the obtained stability margin is just the stability
margin of the corresponding linearized system, which may be inaccurate
when the nonlinearity is strong, or the system state is far away from
the operating point.

\subsection{Example 2: An SISO strongly nonlinear system}

This example aims to verify that the obtained classic frequency-domain
stability margin may be unreliable when the assumption of weak nonlinearity
fails. A strongly nonlinear system is considered 
\begin{align}
\dot{x}_{1} & =x_{1}+x_{2}\nonumber \\
\dot{x}_{2} & =x_{2}+x_{2}^{2}+\mu.\label{Sys2}
\end{align}
Its state-space realization is of the form (\ref{eq:Sys0}) with 
\[
\mathbf{A=}\left[\begin{array}{cc}
1 & 1\\
0 & 1
\end{array}\right],f\left(\mathbf{x}\right)=\left[\begin{array}{c}
0\\
x_{2}^{2}
\end{array}\right],\mathbf{b}=\left[\begin{array}{c}
0\\
1
\end{array}\right].
\]
Notice that $\mathbf{A}$ is unstable, but $\left(\mathbf{A},\mathbf{b}\right)$
is controllable. Thus, a state feedback controller is designed firstly
to obtain a stable $\mathbf{\bar{A}=A-bk}_{0}^{\text{T}}$. Then,
the subsequent stabilizing controller design is similar to that of
\emph{Example 1} and omitted here.

\begin{figure}[ptbh]
\begin{centering}
\includegraphics[width=0.5\textwidth]{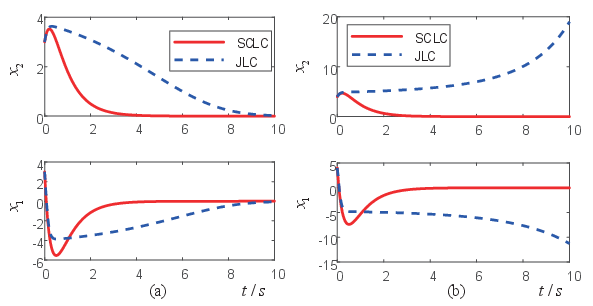}
\par\end{centering}
\caption{State response (Example 3): (a) $\mathbf{x}_{0}=[3$ $3]^{\text{T}}$;
(b) $\mathbf{x}_{0}=[4$ $4]^{\text{T}}$.}
\label{Ex2-1}
\end{figure}

First, choose $\mathbf{k}_{0}=\left[6\text{ }5\right]^{\text{T}}$
to obtain a stable $\mathbf{\bar{A}}$. Then, for a fair comparison,
the parameters in LQR are selected as the same values for the state-compensation-linearization-based
control and JLC. In the following simulations, $\mathbf{Q}=$diag$\left(10\text{,}10\right),R=1$
are selected resulting in $\mathbf{k=-}\left[4.88\text{ }1.98\right]^{\text{T}}$.
Besides, let $c_{1},c_{2}=20$. When the initial state value is set
to $\mathbf{x}_{0}=[3$ $3]^{\text{T}},$ the state response is depicted
in Fig.\ \ref{Ex2-1}(a), which shows that SCLC has a faster state
convergence rate than JLC. Furthermore, when the initial state is
changed to $\mathbf{x}_{0}=[4$ $4]^{\text{T}}$, it can be found
from Fig.\ \ref{Ex2-1}(b) that system divergence occurs for JLC,
while SCLC still performs well.

For JLC, the calculated stability margin is $G_{\text{m}}=\infty$,
$P_{\text{m}}=\infty$. However, the instability phenomenon occurs
when the initial state is $\mathbf{x}_{0}=[4$ $4]^{\text{T}}$. Hence,
it convincingly shows that the classic stability margin is unreliable
in this case. The reason is that the closed-loop system is still nonlinear,
and the calculated stability margin is just valid in a small neighborhood
of the selected operating point (the point where the system is linearized).

\subsection{Example 3: An MIMO nonlinear system}

This example aims to demonstrate the complete procedure to design
the controller and measure the stability margin for a MIMO nonlinear
system by using the proposed method, and to show the difference between
the defined stability margin and the classic frequency-domain stability
margin. Consider a two-input two-output nonlinear system of the form
in (\ref{eq:Sys0}) with

\begin{align}
\mathbf{A} & =\left[\begin{array}{ccc}
-1 & 0 & 1\\
0 & -1 & 1\\
0 & -2 & -3
\end{array}\right],\mathbf{B}=\left[\begin{array}{cc}
0 & -1\\
0 & 1\\
1 & 1
\end{array}\right],\label{eq:Ex1}\\
 & \mathbf{f}\left(\mathbf{x}\right)=\left[\begin{array}{c}
0\\
0\\
x_{3}^{2}\frac{1}{1+ax_{3}^{2}}
\end{array}\right].
\end{align}
where $\mathbf{x}=\left[x_{1}x_{2}x_{3}\right]^{\text{T}}$. It can
be verified that $\mathbf{A}$ is a Hurwitz matrix.

\subsubsection{State-compensation-linearization-based control}

In this part, the controller design and stability margin analysis
will be carried out by following the procedure illustrated in Fig.\,\ref{Fig_Procedure}.

$\bullet$ Step (a). Design controllers for the primary system and
the secondary system. For the controller (\ref{eq:Controller_Prob1-1}),
the LQR method is employed to calculate the feedback matrix $\mathbf{K}$.
Besides, let $\mathbf{H}(s)=\mathbf{I}_{m}$. Then, the primary controller
follows

\[
\mathbf{u}_{\text{p}}=\mathbf{K}\mathbf{\hat{x}_{\text{p}}}.
\]
For the controller (\ref{eq:Controller_Prob2-1}), the Lyapunov method
is adopted, and a secondary controller is designed as

\begin{equation}
\mathbf{u}_{\text{s}}=\left[\begin{array}{c}
-\left(x_{\text{s},1}-x_{\text{s},2}+\left(x_{\text{s},3}+x_{\text{p},3}\right)^{2}\frac{1}{1+a\left(x_{\text{s},3}+x_{\text{p},3}\right)^{2}}\right)\\
-c\left(-x_{\text{s},1}+x_{\text{s},2}+x_{\text{s},3}\right)
\end{array}\right]\label{eq:Us}
\end{equation}
where $c>0$ is the controller parameter to be specified later. The
Lyapunov design procedure is routine, and hence omitted here.

$\bullet$ Step (b). Measure the stability margin for the primary
system to get $\gamma_{\text{max,1}}$ and $\tau_{\text{max,1}}$.
Because the primary system is a MIMO linear system, the proposed $\mathcal{L_{\text{\ensuremath{2}}}}$
stability margin in \cite{https://doi.org/10.1002/rnc.7413} can be
considered.

$\bullet$ Step (c). Measure the stability margin for the whole system
to get $\gamma_{\text{max,2}}$ and $\tau_{\text{max,2}}$. First,
$k_{l}$ is calculated based on (\ref{eq:Us}). Then, based on Section
\ref{Sec:SM OSys}, the frequency-sweep test is carried out by applying
frequency-sweep input signals to the system at the point $\mathbf{q}^{\prime\prime}$
and collecting output responses at the point $\mathbf{p}$, as shown
in Fig.\,\ref{Fig_Procedure}(c). Based on the collected input-output
data, $\left\Vert \mathbf{G}_{0}\mathbf{B}\right\Vert $ and $\left\Vert s\mathbf{G}_{0}\mathbf{B}\right\Vert $
are obtained, and then $\gamma_{\text{max,2}}$ and $\tau_{\text{max,2}}$
are gotten by (\ref{eq:Gamma2}) and (\ref{eq:Tao2}).

\subsubsection{JLC for comparison }

A commonly-used way is to apply linear control techniques at discrete
operating points within the operating envelope, and then classic frequency-domain
stability margin is used to evaluate the designed controllers at the
discrete points. For stabilizing control, the operating point is the
equilibrium point, i.e., the origin herein. By linearizing the system
(\ref{eq:Ex1}) at the origin $\mathbf{x}=\mathbf{0}$, one has 
\begin{align}
\dot{\mathbf{x}}=\mathbf{Ax}+\mathbf{Bu.}\label{eq:LSys}
\end{align}
Then, similar to the controller (\ref{eq:Controller_Prob1-1}), an
LQR-based stabilizing controller for (\ref{eq:LSys}) is designed
as

\begin{equation}
\mathbf{u}=\mathbf{K}\mathbf{x}.\label{eq:DLBC1}
\end{equation}
Through neglecting the coupling among loops, the stability margin
analysis for the closed-loop system consisting of (\ref{eq:LSys})
and (\ref{eq:DLBC1}) can be done by classic frequency-domain stability
margin. In practice, the Bode plot is obtained by the frequency-sweep
technique, by which the resulting stability margin is obtained.

\subsubsection{Simulation results}

For a fair comparison, the parameters in LQR are selected as the same
values for state-compensation-linearization-based control and JLC.
In the following simulations, $\mathbf{Q}=\text{diag\ensuremath{\left(1,1,1\right)}},\mathbf{R}=\text{diag}\left(1,1\right)$
are selected. Besides, let $a=0.01,\mathbf{x_{0}}=\left[10\:10\:10\right]^{\text{T}},c=5$.

Stability margin analysis is performed after the stabilizing control
has been done. For state-compensation-linearization-based control,
the stability margin for the primary system is investigated firstly.
The stability margin is obtained in two ways: the theoretical computation
method and the frequency-sweep method. Through theoretical computation,
the stability margin for the primary system is obtained as $\gamma_{\text{max,1}}=2.262$
and $\tau_{\text{max,1}}=1.134$, while through the frequency-sweep
method, it is acquired that $\gamma_{\text{max,1}}=2.261$ and $\tau_{\text{max,1}}=1.134$.
Thus, the primary system is stable, i.e., $\mathbf{G}_{\Delta}:\mathcal{L}_{\text{\ensuremath{2}}}^{\text{\ensuremath{m}}}\rightarrow\mathcal{L}_{\text{\ensuremath{2}}}^{\text{\ensuremath{n}}}$.

Then, the stability margin for the whole system is also obtained in
the mentioned two ways. Through theoretical computation, the stability
margin for the whole system is obtained as $\gamma_{\text{max,2}}=0.17$
and $\tau_{\text{max,2}}=0.08$, while through the frequency-sweep
method, it is obtained that $\gamma_{\text{max,2}}=0.19$ and $\tau_{\text{max,2}}=0.08$.
Note that the two groups of results show a good agreement with each
other. Finally, $\gamma_{\text{max}}\thickapprox\min\left(\gamma_{\text{max,1}},\gamma_{\text{max,2}}\right)=0.19$,
$\tau_{\text{max}}\thickapprox\min\left(\tau_{\text{max,1}},\tau_{\text{max,2}}\right)=0.08$.

\begin{figure}
\begin{centering}
\includegraphics[width=0.5\textwidth]{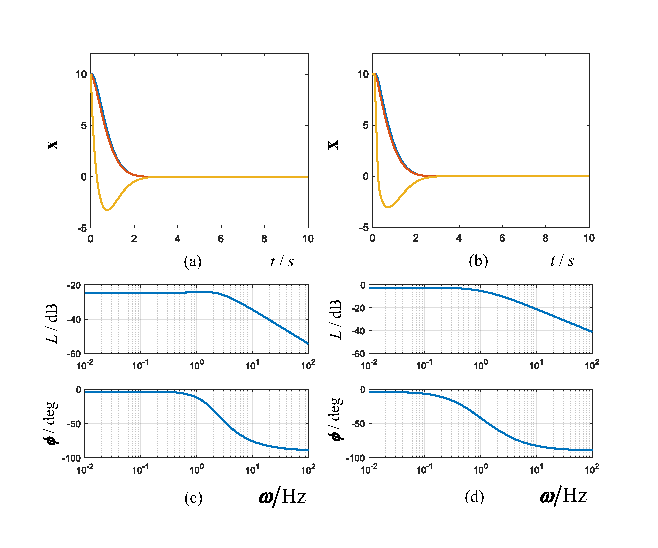}
\par\end{centering}
\caption{stability margin analysis results (Example 4): (a) Validation test
results of $\gamma_{\text{max}}$; (b) Validation test results of
$\tau_{\text{max}}$; (c) Bode plot of the first loop in JLC; (d)
Bode plot of the second loop in JLC.}
\label{Fig_Ex1}
\end{figure}

To verify the obtained $\gamma_{\text{max}}$, $\Delta=\text{diag\ensuremath{\left(0.19,0.19\right)} }$
is added to the original system, and the corresponding state response
is depicted in Fig.\,\ref{Fig_Ex1}(a). To verify the obtained $\tau_{\text{max}}$,
$\Delta=\textrm{diag}\left(e^{-0.08s}-1,e^{-0.08s}-1\right)$ is added
to the original system, and the corresponding state response is depicted
in Fig.\,\ref{Fig_Ex1}(b). It can be observed that the system is
still stable.

For JLC, the current practice is to certify such systems by the GM
and PM at the origin. There are two loops for the considered system.
From Fig.\,\ref{Fig_Ex1}(c),(d), the GM and PM for the two loops
are $G_{\text{m}}=\infty,P_{\text{m}}=\infty$. Actually, the method
directly treats the nonlinear system as a linear one, and the obtained
stability margin is just the stability margin of the corresponding
linearized system. Moreover, the coupling between the two loops is
neglected, and each loop is considered individually. The obtained
stability margin may be inaccurate when coupling or nonlinearity is
strong, or the system state is far away from the operating point.

\subsection{Discussions}

From the simulation results of the three illustrative examples, it
can be seen that the state-compensation-linearization-based stabilizing
control and stability margin analysis outperforms the traditional
stabilizing control and stability margin analysis in two aspects.
First, compared with JLC and FLC, the stabilizing control performance
of state-compensation-linearization-based control is better. State-compensation-linearization-based
control can achieve a higher state convergence rate and avoid the
uncontrollability of JLC and the singularity of FLC. Secondly, the
subsequent stability margin analysis can make the classical frequency-domain
stability margin available for nonlinear systems. The classical frequency-domain
stability margin may be insufficient when it is applied to nonlinear
systems directly owing to the nonlinearity. However, it can work well
under the proposed state-compensation-linearization-based control
framework by utilizing the small-gain theorem. What is more, a frequency-domain
series compensator, namely $\mathbf{H}\left(s\right)$ in the controller
(\ref{eq:Controller_Prob1-1}) can be introduced additionally to pursue
a better stability margin when the actual stability margin is unsatisfactory.

\section{Conclusions}

\label{Sec:Conclusions}

Based on the state-compensation-linearization-based \textit{\emph{stabilizing
control}}, the paper focuses on the stability margin analysis for
nonlinear systems. The $\mathcal{L}_{2}$ gain margin and $\mathcal{L}_{2}$
time-delay margin for the closed-loop nonlinear system are defined,
and derived by the small-gain theorem. The stability margin measurement
depends on the frequency-sweep method. Three examples are provided
to show the concrete design and analysis procedure. Simulation results
illustrate that the state-compensation-linearization-based control
outperforms Jacobian linearization based control and feedback linearization
based control to some extent. It should be noted that the frequency-domain
stability margin analysis is extended smoothly to nonlinear systems.
As a result, it is supposed to be of interest to many engineers.

\bibliographystyle{IEEEtranTIE}
\bibliography{IEEEabrv}

\end{document}